\documentclass[aps,showpacs,nofootinbib,amsmath,amssymb,12pt]{revtex4-1}
\usepackage{graphicx}

\begin{document}

\title{
A simple test for stability of black hole by $S$-deformation
}

\author{
Masashi Kimura
}

\affiliation{
CENTRA, Departamento de F\'{\i}sica, Instituto Superior T\'ecnico, Universidade de Lisboa, Avenida~Rovisco Pais 1, 1049 Lisboa, Portugal
}

\date{\today}


\begin{abstract}
We study a sufficient condition to prove the stability of
a black hole when the master equation for linear perturbation
takes the form of the Schr\"odinger equation.
If the potential contains a small negative region,
usually, the $S$-deformation method was used to show
the non-existence of unstable mode.
However, in some cases, it is hard to
find an appropriate deformation function analytically
because the only way known so far to find it is a try-and-error approach.
In this paper, we show that it is easy to find a regular deformation function
by numerically solving the differential equation
such that the deformed potential vanishes everywhere, when the spacetime is stable.
Even if the spacetime is almost marginally stable, our method still works.
We also discuss a simple toy model which can be solved analytically,
and show the condition for the non-existence of a bound state is
the same as that for the existence of a regular solution for the differential equation in our method.
{}From these results, we conjecture that our criteria is also a necessary condition.
\end{abstract}

\maketitle

\section{Introduction}
It is known that the equations for gravitational perturbation
around a highly symmetric black hole spacetime
usually reduce to decoupled master equations~\cite{Regge:1957td, Kodama:2003jz, Vishveshwara:1970cc, Zerilli:1970se, Zerilli:1971wd, Ishibashi:2003ap, Kodama:2003kk, Dotti:2004sh, Dotti:2005sq, Gleiser:2005ra, Takahashi:2010ye, Takahashi:2009dz, Takahashi:2009xh, Kimura:2007cr, Nishikawa:2010zg} in the form
\begin{align}
\left[
-\frac{\partial^2 }{\partial t^2}  +\frac{\partial^2 }{\partial x^2}  - V(x) \right]\tilde{\Phi} = 0.
\label{mastereqtr}
\end{align}
Using the Fourier transformation with respect to time coordinate, 
$\tilde{\Phi}(t,x) = e^{-i\omega t}\Phi(x)$, the master equation
takes the form of the Schr\"odinger equation 
\begin{align}
\left[
-\frac{d^2}{dx^2} + V\right]\Phi = \omega^2 \Phi.
\label{schrodingereq}
\end{align}
When we wish to prove the stability of the black hole spacetime,
we need to show the non-existence of $\omega^2 < 0$ solution, {\it i.e.,} non-existence of 
exponentially growing solution,
under the boundary conditions,\footnote{
In this paper, since we mainly focus on asymptotically flat (or de Sitter) black holes,
we assume that the range of $x$, which is the tortoise coordinate, is $-\infty < x < \infty$.
Here, $x \to -\infty$ and $x \to \infty$ correspond to the black hole horizon and the spatial infinity 
(or cosmological horizon), respectively.
If we consider asymptotically anti-de Sitter black holes,
the range of $x$ becomes $-\infty < x < x_{\rm max}$ with the finite value $x_{\rm max}$.
}
$\Phi\to 0, d\Phi/dx \to 0$ at $x \to \pm \infty$, and the conditions: $\Phi$
and $d\Phi/dx$ are continuous and bounded everywhere.
{}From Eq.\eqref{schrodingereq}, we obtain
\begin{align}
-
\left[
\bar{\Phi}\frac{d\Phi}{dx} 
\right]_{- \infty}^{\infty}
+
\int dx
\left[
\left|\frac{d\Phi}{dx}\right|^2 + V  \left|\Phi \right|^2 
\right]
=
\omega^2 \int dx\left|\Phi \right|^2,
\end{align}
where $\bar{\Phi}$ is the complex conjugate of $\Phi$. 
The first term in LHS vanishes from the above boundary conditions.
If the effective potential $V$ is non-negative everywhere, this implies 
$\omega^2 \ge 0$.

In some cases, even if the effective potential contains a small negative region, the spacetime still can be shown to be stable against linear perturbation in a following way.
{}From Eq.\eqref{schrodingereq}, we can also show
\begin{align}
-\frac{d}{dx}
\left[
\bar{\Phi}\frac{d\Phi}{dx}  + S |\Phi|^2
\right]
+
\left|\frac{d\Phi}{dx} + S \Phi \right|^2 + \left(V + \frac{dS}{dx} - S^2 \right) \left|\Phi \right|^2 
=
\omega^2 \left|\Phi \right|^2,
\label{sdef1}
\end{align}
where $S$ is an arbitrary function of $x$.
We impose continuity on $S$ everywhere, then
the integral
\begin{align}
\int dx \frac{d}{dx}(S |\Phi|^2),
\label{surfaceterms}
\end{align}
becomes a surface term. 
Integrating Eq.\eqref{sdef1}, we obtain
\begin{align}
-
\left[
\bar{\Phi}\frac{d \Phi}{dx} + S |\Phi|^2
\right]_{- \infty}^{\infty}
+
\int dx
\left[
\left|\frac{d\Phi}{dx} + S \Phi \right|^2 + \left(V + \frac{dS}{dx} - S^2 \right) \left|\Phi \right|^2 
\right]
=
\omega^2 \int dx\left|\Phi \right|^2,
\end{align}
Also, we assume that $S$ is not divergent at $x \to \pm \infty$ so that the above surface term vanishes only
from the boundary condition for $\Phi$.
The potential term is deformed as
\begin{align}
\tilde{V} := V + \frac{dS}{dx} - S^2.
\label{sdeformation}
\end{align}
This is called the $S$-deformation of the potential $V$.
If the deformed effective potential $\tilde{V}$
is non-negative everywhere by choosing an appropriate function $S$,
we can also say $\omega^2 \ge 0$.
In~\cite{Kodama:2003jz, Ishibashi:2003ap, Kodama:2003kk, Dotti:2004sh, Dotti:2005sq, Gleiser:2005ra, Beroiz:2007gp,  Takahashi:2009xh, Takahashi:2010gz}, the stability of spacetime was shown analytically by using this $S$-deformation method.
However, the only way known so far 
to find an appropriate function $S$ analytically is a try-and-error approach.
When it is hard to find an appropriate $S$ deformation, a numerical study is needed.
In~\cite{Konoplya:2007jv, Konoplya:2008ix, Ishihara:2008re, Konoplya:2008yy, Konoplya:2008au, Konoplya:2013sba},
stability and instability were investigated
by numerically solving the two dimensional partial differential equation~\eqref{mastereqtr} in time domain.

In this paper, we
propose a simple way to show the stability by showing
the existence of the $S$-deformation such that the deformed potential $\tilde{V}$ vanishes.
For this purpose, we need to solve the equation
\begin{align}
V + \frac{dS}{dx} - S^2 = 0.
\label{sdefvzero}
\end{align}
The approximate solution near $x \to \pm \infty$ 
is $S \simeq -1/(c_\pm -x)$ with constants $c_\pm$ if the potential rapidly decays to zero.
Since Eq.\eqref{sdefvzero} is a first order ordinary differential equation,
all solutions near $x \to \pm \infty$ should 
behave this approximate solution.
To obtain a solution with finite $S$ at $x \to \pm \infty$,
we need to find a boundary condition so that
$S$ is positive near $x \to -\infty$ and $S$ is negative near $x \to \infty$,
then $S$ behaves $-1/x$ and finite at $x \to \pm \infty$.
While $S$ diverges at some point for an inappropriate boundary condition, 
we can find the continuous range of appropriate boundary conditions, which corresponds to bounded $S$, 
for typical examples.

One may think that to solve the equation $V + dS/dx - S^2 = W$ for a positive function 
$W (> 0)$ is easier
than to solve Eq.\eqref{sdefvzeromain}. However, since this equation can be written in the form 
$(V - W) + dS/dx - S^2 = 0$, the problem is to find a solution of Eq.\eqref{sdefvzeromain}
for a deeper potential $V - W$, {\it i.e.,} more dangerous case.
This suggests that to solve Eq.\eqref{sdefvzeromain} is the most efficient way to find an appropriate function $S$.

This paper is organized as follows. In the next section,
we discuss the existence of the solution of Eq.\eqref{sdefvzero} in the case of positive potential,
and that in a simple toy model which can be solved analytically.
We also discuss the relation between the solution of 
Eq.\eqref{sdefvzero} in marginally stable case
and the solution of Eq.\eqref{schrodingereq} with $\omega^2 = 0$ .
In Sec.\ref{seciii}, we numerically solve Eq.\eqref{sdefvzero} for 
the higher-dimensional spherically-symmetric black holes, and construct appropriate 
deformation functions from various boundary conditions.
Sec.\ref{summary} is devoted to summary and discussion.

\section{On existence of solution}
\label{secii}

\subsection{Local existence}
Let us consider
the differential equation 
\begin{align}
V(x) + \frac{dS(x)}{dx} - S(x)^2 = 0.
\label{sdefvzeromain}
\end{align}
We assume that $V$ is continuous and bounded in $-\infty < x < \infty$.
From the uniqueness theorem of the ordinary differential equation, 
there exists a solution of the above equation, at least, locally.

We can see this by considering the Taylor expansion if $S$ and $V$ are analytic functions.
The series around some point $x = x_0$ are
\begin{align}
S = \sum_{n = 0}^\infty s_n (x-x_0)^n,~~
V = \sum_{n = 0}^\infty v_n (x-x_0)^n,
\end{align}
{}From the differential equation, we obtain the coefficients of $S$ as
\begin{align}
s_{n + 1} = \frac{1}{n+1}\left[
-v_n + \sum_{m = 0}^n s_m s_{n-m}
\right],
\end{align}
where $s_0$ is an arbitrary constant  which corresponds to the integration constant (or boundary condition).
This shows that we can find a solution $S$ locally.

In general, while $dS/dx - S^2$ is finite, $S$ might be divergent at a finite coordinate value $x$.
The problem is to find a function $S$ which is continuous and bounded everywhere.
In some cases, we can show the existence of such the regular function $S$.

\subsection{Positive potential}
\label{positivepotentialcase}

First, we consider that the potential is positive and bounded above in $-\infty < x < \infty$.
While this corresponds to the manifestly stable case, to show the existence of 
a continuous and bounded solution of Eq.\eqref{sdefvzeromain} is not trivial.
We would like to show the following proposition.
\\\\
{\bf Proposition 1.}~~{\it If the potential $V$ is positive and bounded above in $-\infty < x < \infty$, 
there exists a continuous and bounded solution of Eq.\eqref{sdefvzeromain} in $-\infty < x < \infty$.
}

\begin{figure}
\includegraphics[width=7cm]{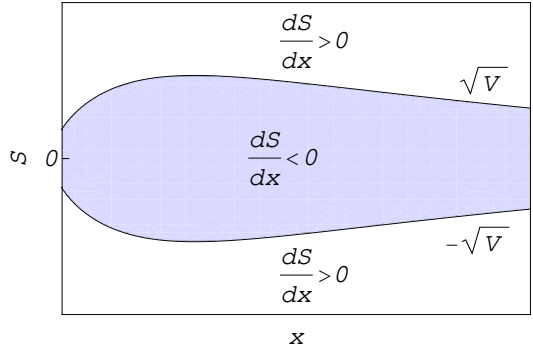}
\caption{\label{fig0} 
The relation among $\pm \sqrt{V}$, the value of $S$ and the sign of $dS/dx$ in $V > 0$ region.
$dS/dx < 0$ if $S$ is in $ -\sqrt{V} < S < \sqrt{V}$ (shaded region),
and
 $dS/dx > 0$ if $S$ is in $S < \sqrt{V}$ or $\sqrt{V} < S$ (white region).
 }
\end{figure}
The proof is given in Appendix.\ref{proof1}.
We can understand this proposition as follows:
Since we have $dS/dx = (S - \sqrt{V})(S + \sqrt{V})$, 
the relation among $\pm \sqrt{V}$, the value of $S$ and the sign of $dS/dx$  becomes like Fig.\ref{fig0}.
If we choose the value of $S$ 
in $-\sqrt{V}|_{x = x_0} < S|_{x = x_0} < \sqrt{V}|_{x = x_0}$ at some point $x = x_0$
as a boundary condition, 
we can say that the solution of Eq.\eqref{sdefvzeromain} satisfies 
$ -\sqrt{V_{\rm max}} < S < \sqrt{V_{\rm max}}$.
Thus, $S$ is bounded above and below in the region $-\infty < x < \infty$.

\subsection{Toy model}
\label{toymodel}
\begin{figure}
\includegraphics[width=7cm]{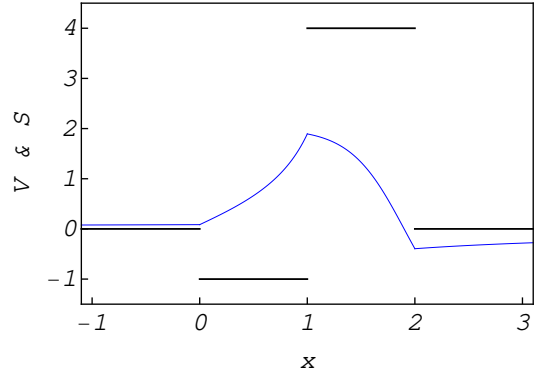}
\caption{\label{figtoymodel}
The effective potential $V$ (thick black) and the typical solution $S$ (solid blue) for 
the parameter of the potential 
$x_1 = 0, x_2 = 1, x_3 = 2, h_1 = 1, h_2 = 2$ and 
the boundary condition for $S$ as $S|_{x = 1.9} = 0$.}
\end{figure}

To obtain a qualitative understanding, 
we consider a toy model of the potential, which can be solved analytically,
\begin{align}
V = 
\begin{cases}
 0   &(-\infty < x \le x_1)
\\ 
- h_1^2 ~( < 0)   &(x_1 < x \le x_2)
\\
 h_2^2 ~( > 0) &(x_2 < x \le x_3)
\\
  0  & (x_3 < x <  \infty)
\end{cases}
\end{align}
with constants $h_1 >0, h_2> 0$.
The local solutions of Eq.\eqref{sdefvzeromain} are\footnote{
Defining $\alpha := e^{2 c_3}$, then $S$ in $x_2 < x \le x_3$ becomes 
$S = -h_2 (-\alpha + e^{2h_2-x})/(\alpha + e^{2h_2 -x})$.
We can see that $\alpha < 0$ case corresponds a complex value of $c_3$,
and $S$ might be divergent in that case. 
The solutions with $\alpha < 0$ satisfy $S^2 > h_2^2$.
To construct continuous $S$ in this subsection, we need to consider $S^2 < h_2^2$ ({\it i.e.,} real $c_3$)
so that $S$ can take both negative and positive values in $x_2 < x \le x_3$.}
\begin{align}
S = 
\begin{cases}
 \dfrac{1}{c_1 - x},   &(-\infty < x \le x_1)
\\ 
h_1 \tan(h_1 x + c_2),   &(x_1 < x \le x_2)
\\
- h_2 \tanh(h_2 x + c_3), &(x_2 < x \le x_3)
\\
 \dfrac{1}{c_4 - x},   & (x_3 < x < \infty)
\end{cases}
\end{align}
where $c_1, c_2, c_3, c_4$ are integration constants.
From the continuity of $S$ at $x = x_1, x_2, x_3$, we obtain the conditions
\begin{align}
&  \frac{1}{c_1 - x_1} - h_1 \tan(c_2 + h_1 x_1) = 0,
\\
& h_1 \tan(c_2 + h_1 x_2) + h_2 \tanh(c_3 + h_2 x_2) =0,
\label{match02}
\\
& \frac{1}{-c_4 + x_3} - h_2 \tanh(c_3 + h_2 x_3) = 0.
\end{align}
From the conditions $S|_{x = x_1} \ge 0, S|_{x = x_2} >0, S|_{x = x_3} \le 0$ 
and the finiteness of $S$,
we obtain the inequalities
\begin{align}
& 
0 <   c_1 - x_1,
~~ 
0 \le  h_1 x_1 + c_2 < \frac{\pi}{2},
\\
& 
0 <  h_1 x_2 + c_2 < \frac{\pi}{2},
~~
h_2 x_2 + c_3 < 0,
\\
&   
0 \le h_2 x_3 + c_3,
~~
c_4 - x_3 < 0.
\end{align}
If the above equations and inequalities are satisfied, then $S$ is continuous and bounded everywhere.
We plot the typical profile of $V$ and $S$ in Fig.\ref{figtoymodel}.
Note that the derivative of $S$ is not necessary to be continuous,
because we only impose the condition such that 
Eq.\eqref{surfaceterms} becomes a surface term.

We would like to derive an inequality between the areas of the potential in negative and positive regions.
{}From the above matching conditions and inequalities, we can show\footnote{
The derivation is as follows:
\begin{align}
 h_1 \tan(h_1 (x_2 - x_1)) &\le h_1 \tan(c_2 + h_1 x_1 + h_1 (x_2 - x_1)) ~~(\because 0 \le c_2 + h_1 x_1)
\notag\\&= 
h_1 \tan(c_2 +  h_1 x_2)
\notag\\&= 
- h_2 \tanh(c_3 + h_2 x_2) ~~(\because {\rm Eq.}\eqref{match02})
\notag\\&=
 h_2 \tanh(-c_3 - h_2 x_2) 
\notag\\&\le
 h_2 \tanh(h_2 x_3- h_2 x_2). ~~(\because 0 \le h_2 x_3 + c_3)
\notag
\end{align}
}
\begin{align}
 h_1 \tan(h_1 (x_2 - x_1))  \le h_2 \tanh(h_2 (x_3- x_2) )
\end{align}
Since we have $X < \tan X $ for $0 < X < \pi/2$, and $\tanh Y < Y$ for $0 < Y$,
we obtain an inequality
\begin{align}
 h_1^2 (x_2 - x_1) < h_2^2 (x_3- x_2).
\end{align}
The area of the negative region is smaller than that of the positive region.
This is consistent with the result in~\cite{Buell:1995} where the existence of an unstable mode 
is shown if $\int_{-\infty}^{\infty} dx V < 0$.

\subsection{Relation between existence of $S$ and non-existence of bound state in toy model}
To check the relation between the existence of regular $S$ and the non-existence of bound state,
we further study the toy model in the previous subsection and 
the dependence of their existence on the ratio of the areas
\begin{align}
\Gamma = \frac{h_2^2(x_3 - x_2)}{h_1^2(x_2 - x_1)}.
\end{align}
First, for simplicity, we consider the case
\begin{align}
x_1 = 0,~ x_2 = 1,~x_3 = 2,~h_1 = 1.
\end{align}
If $\Gamma = 0$, this is just a single well problem.
In that case, there is only a bound state in the solution of the Schr\"odinger equation 
whose energy is $\omega^2|_{\Gamma = 0} \simeq -0.43$.
After some calculation, we obtain the condition for the existence of the bound state, 
which can be calculated from $\omega^2 < 0$,
as
\begin{align}
\Gamma < \Gamma_{\rm cr}
\end{align}
where $\Gamma_{\rm cr} \simeq 2.79$ is defined by 
\begin{align}
\tan(1) = \sqrt{\Gamma_{\rm cr}}\tanh(\sqrt{\Gamma_{\rm cr}}).
\label{criticalgamma}
\end{align}
Also, we can calculate the condition for the existence of continuous and bounded solution $S$ of Eq.\eqref{sdefvzeromain}.
That condition, which is derived from $S|_{x = x_1} \ge 0$ with $S|_{x = x_3} = 0$, 
becomes $\Gamma \ge \Gamma_{\rm cr}$ 
with the same critical value in Eq.\eqref{criticalgamma}.

For general parameters, we can derive the same relations.
The critical value $\Gamma_{\rm cr}$ in general case is defined by~\footnote{From
$X < \tan X $ for $0 < X < \pi/2$, and $\tanh Y < Y$ for $0 < Y$, we can see $\Gamma_{\rm cr} >1$.
If the arguments of both tangent and hyperbolic tangent
are small, $\Gamma_{\rm cr} \simeq 1$.}
\begin{align}
\frac{\sqrt{x_3-x_2}}{\sqrt{x_2-x_1}}\tan(h_1(x_2-x_1)) = \sqrt{\Gamma_{\rm cr}}
\tanh(h_1\sqrt{\Gamma_{\rm cr}}\sqrt{x_2-x_1}\sqrt{x_3-x_2}).
\label{criticalgammageneral}
\end{align}
Note that if $h_1(x_2 - x_1) \ge \pi/2$, there exists at least one bound state regardless of 
the value of $h_2$, in that case, $\Gamma_{\rm cr}$ is not defined.
Thus, in this toy model, the condition for
the non-existence of bound state 
coincides with that for the existence of continuous and bounded solution $S$.

\subsection{Relation between regular $S$ in marginally stable case and onset of unstable mode}
\label{marginalystable}

Since Eq.\eqref{sdefvzeromain} is the Riccati equation, defining 
\begin{align}
\frac{1}{\phi}\frac{d\phi}{dx} := - S,
\label{riccatieq}
\end{align}
we can write Eq.\eqref{sdefvzeromain} in a second order linear ordinary differential equation
\begin{align}
- \frac{d^2\phi}{dx^2} + V \phi = 0.
\label{zeroenergysch}
\end{align}
This is the master equation \eqref{schrodingereq} with $\omega^2 = 0$.
If this equation has a non-trivial regular solution, 
it probably corresponds to the onset of unstable mode.\footnote{
We should note that we cannot construct a regular solution of 
Eq.\eqref{zeroenergysch} from a regular solution of 
Eq.\eqref{sdefvzeromain} except for the marginally stable case.
This is because the asymptotic behavior of the regular solution of 
Eq.\eqref{sdefvzeromain} is $S \sim -1/x$, but the corresponding $\phi$
behaves $\phi \sim c x$ with a constant $c$, which is not a regular solution of
Eq.\eqref{zeroenergysch}.
}
We can expect that there is some relation between the solutions of 
Eq.\eqref{sdefvzeromain} in marginally stable case 
and the onset of unstable mode.
In this subsection, we briefly discuss this.

Suppose that the potential contains a parameter $\alpha$,
and the master equation has a property such that 
there is no unstable mode if $\alpha \le \alpha_{\rm cr}$ 
and there are unstable modes if $\alpha > \alpha_{\rm cr}$.
In the case of $\alpha > \alpha_{\rm cr}$, we define $\Phi_0(x;\omega(\alpha))$
as the ground state of the Sch\"odinger equation~\eqref{schrodingereq},
then the energy takes negative value $\omega^2 < 0$.
If we assume that the potential rapidly decays in $x \to \pm \infty$,
then $\Phi_0 \sim e^{\mp x \sqrt{-\omega^2}}$ at $x \to \pm \infty$.
In that case, $\Phi_0^{-1} d\Phi_0/dx$ takes finite values at the boundary $x \to \pm \infty$.
Since it is known that the ground state for the one-dimensional Sch\"odinger equation has 
no nodes (for example, see~\cite{MessiahvolI, Moriconi:2007}), $\Phi_0$ cannot becomes zero except at the boundary.
So, we can define a function
\begin{align}
{\cal S} := - \frac{1}{\Phi_0(x;\omega(\alpha))}\frac{d\Phi_0(x;\omega(\alpha))}{dx},
\end{align}
and this is regular in $-\infty < x < \infty$.
We can show that ${\cal S}$ satisfies
\begin{align}
V + \frac{d {\cal S}}{dx} - {\cal S}^2 = \omega^2.
\end{align}
If we assume the function ${\cal S}$ has a smooth limit in $\alpha \to \alpha_{\rm cr} +0$,
this becomes a regular solution of Eq.\eqref{sdefvzeromain}
since $\omega^2 \to 0$ in this limit.

In fact, the toy model in the previous subsections satisfies the above assumptions.
This is the reason why the critical values are the same in the toy model.
In general, the validity of the above assumption is not clear, but it seems to be reasonable.

\section{numerical calculation}
\label{seciii}

If the effective potential $V$ rapidly decays as $x \to \pm \infty$,
the approximate solution of Eq.\eqref{sdefvzeromain} becomes $-1/(c_\pm - x)$ with constants $c_\pm$.
To obtain a solution with finite $S$ at $x \to \pm \infty$,
we need to find a boundary condition so that
$S$ is positive near $x \to -\infty$ and $S$ is negative near $x \to \infty$.
Then, $S$ behaves $-1/x$ and finite at $x \to \pm \infty$.
A bounded $S$ usually changes the value from positive to negative 
at some point in $V > 0$ region so that $S$ takes negative value near $x \to \infty$ 
like $S$ in Fig.\ref{figtoymodel}.
We can expect that 
the appropriate boundary conditions can be found by setting $S = 0$ at a point where $V$ is positive.
In this section, we study the stability of 10-dimensional spherically symmetric black holes
and the 5-dimensional Schwarzschild black string, as typical examples.
The explicit form of the metrics and the master equations, 
which was derived in~\cite{Kodama:2003jz, Ishibashi:2003ap, Kodama:2003kk, Gregory:1993vy, Hovdebo:2006jy, Konoplya:2008yy}, is given in Appendices.\ref{appendixa} and \ref{appendixblackstring}.

In the following examples, we used the function {\tt NDSolve} in {\it Mathematica} 
to solve the equation numerically.
In a stable case, for an appropriate boundary condition, 
we could obtain a bounded function $S$ without a special technic,
as far as we confirmed.
We estimated the numerical error by $\epsilon := (V + dS/dx - S^2)/(|V| + |dS/dx| + |S^2|)$
and confirmed $\epsilon < 10^{-6}$ in the following calculations.

\subsection{10-Dimensional Schwarzschild black hole}
\begin{figure}
\includegraphics[width=9cm]{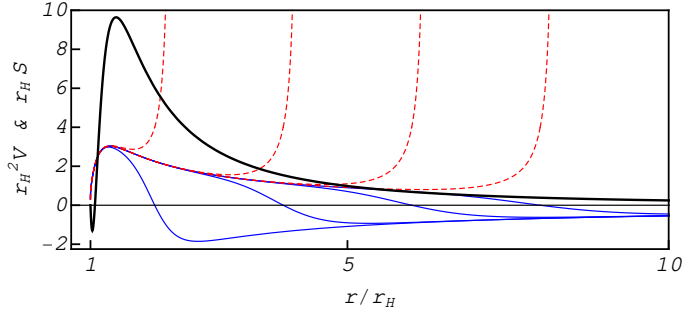}
\caption{\label{fig1} The effective potential $V$
of the $\ell = 2$ vector perturbation in the 10-Dimensional Schwarzschild black hole (thick black),
and the numerical solution $S$ for various boundary conditions (solid blue and dashed red).
Solid blue lines correspond to bounded $S$, solid red lines correspond to unbounded $S$.
The boundary conditions for blue lines are $S = 0$ at $r/r_H =2, 4, 6, 8$, respectively,
and that for red lines are $r_H S = 4$  at $r/r_H =2 , 4, 6, 8$, respectively.
 }
\end{figure}
Let us consider the $\ell = 2$ vector perturbation in the 10-Dimensional Schwarzschild black hole.
In fact, in this case, the existence of the $S$-deformation such that the deformed potential is positive 
is already known~\cite{Ishibashi:2003ap}, but this is still a good example to check that our method works well.
In Fig.\ref{fig1}, we plot the effective potential, which contains a small negative region near the horizon, 
and the numerical solution of Eq.\eqref{sdefvzeromain}
for various boundary conditions. 
Note that we use the radial coordinate $r$ in the black hole spacetime, 
the relation between $r$ and $x$ is in Appendix.\ref{appendixa}.
If we plot $-1/x$ as a function of $r$, 
it seems to be rapidly decrease to zero from a finite value
near the horizon since $-1/x \sim -1/\ln(r-r_H)$.
There are the two attractors of solutions in $V> 0$ region, and they are almost $\pm \sqrt{V}$.
This is because $dS/dx = (S + \sqrt{V})(S - \sqrt{V})$ in $V>0$ region (see also Fig.\ref{fig0}).
We can see that the bounded solutions can be found by choosing the 
boundary condition as $S = 0$ at the points where $V > 0$ (solid blue lines in Fig.\ref{fig1}).
If we choose the boundary condition of $S$ larger than $\sqrt{V}$ in $V > 0$ region,
then $S$ is divergent at some point (dashed red lines in Fig.\ref{fig1}).

Unlike the toy model in Sec.\ref{toymodel},
the potential is negative near the horizon.
Since the potential behaves $V \simeq -a^2 (r - r_H) \sim - a^2e^{b^2 x}$ with constants $a>0, b>0$ 
near the horizon, it rapidly decays in $x$ coordinate.
In that case, if the $S$ takes positive value near the horizon, 
$S$ will be bounded above and below as $x$ decreases (see Appendix.\ref{appendixb}).

\subsection{10-Dimensional Schwarzschild-de Sitter black hole}
\begin{figure}
\includegraphics[width=9cm]{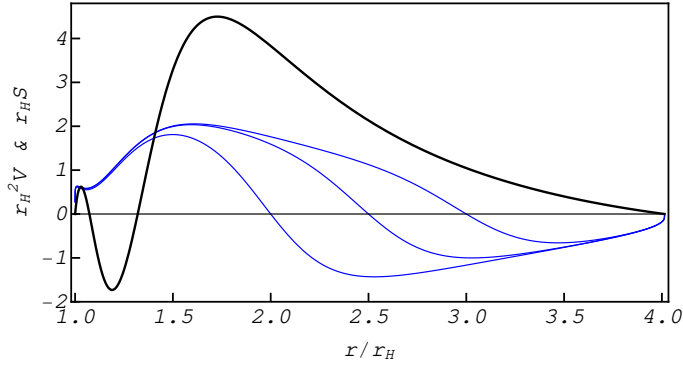}
\caption{\label{fig2}  The effective potential $V$
of the $\ell = 2$ scalar perturbation in the 10-Dimensional Schwarzschild-de Sitter black hole (thick black)
with $r_H^2 \lambda  = 0.05$,
and the numerical solution $S$ for various boundary conditions (solid blue).
Solid blue lines correspond to bounded $S$ and 
the boundary conditions are $S = 0$ at $r/r_H =2, 2.5, 3$, respectively.
Here, we only plot bounded solutions.
 }
\end{figure}
As an another example, we consider the $\ell = 2$ scalar perturbation in the 10-Dimensional Schwarzschild-de Sitter black hole. 
In this case, the existence of the $S$-deformation such that the deformed potential is positive 
is not known, but there is a numerical proof of stability based on the quasi normal mode~\cite{Konoplya:2007jv}.
In Fig.\ref{fig2}, we plot the effective potential and the numerical solution of Eq.\eqref{sdefvzeromain}
for various boundary conditions when the cosmological constant is $r_H^2 \lambda  = 0.05$. The effective potential also contains a small negative region near the horizon.
We can see that the solutions are bounded above and below
if we choose boundary condition as $S = 0$ at the points where $V > 0$ (blue lines in Fig.\ref{fig1}).
Since our method is different from the previous work~\cite{Konoplya:2007jv}, 
this is a complementary result which also supports the stability of the spacetime.

\subsection{10-Dimensional Reissner-Nordstr\"om-de Sitter black hole}
\label{10RNbh}
\begin{figure}
\includegraphics[width=8cm]{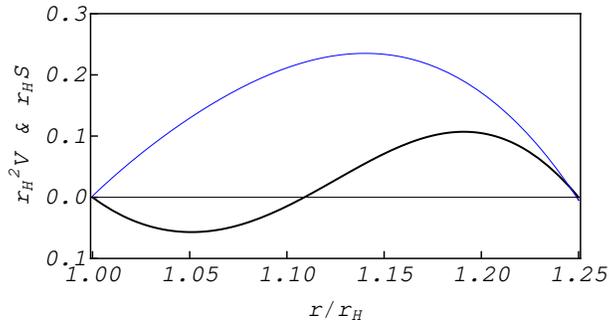}
\caption{\label{fig3}  The effective potential $V$
of the $\ell = 2$ scalar gravitational perturbation in the 10-Dimensional Reissner-Nordstr\"om-de Sitter black hole 
(thick black)
with the parameter $r_{dS}/r_{H} = 5/4, Q/Q_{\rm extremal} = 0.74$,
and the bounded solution $S$ with the boundary condition $S|_{r = r_{dS} - 10^{-3}r_{H}} = 0$ (solid blue).
 }
\end{figure}
We consider $\ell = 2$ scalar gravitational perturbation in 
the 10-Dimensional Reissner-Nordstr\"om-de Sitter black hole.
In this case, it is known that 
there exists unstable mode for 
large values of electric charge 
and cosmological constant~\cite{Konoplya:2008au, Cardoso:2010rz, Konoplya:2013sba}.
According to Fig.4 in~\cite{Konoplya:2008au}, for the parameter $r_H =1, r_{dS} = 5/4$,
if the electric charge is larger than the critical value around $Q/Q_{\rm extremal} \simeq 0.75$,
the spacetime is unstable.
In Fig.\ref{fig3}, we plot the effective potential and the numerical solution of Eq.\eqref{sdefvzeromain}
for $r_H =1, r_{dS} = 5/4, Q/Q_{\rm extremal} = 0.74$.
We can find a continuous and bounded $S$ which is positive at $r \simeq r_H$ and negative at $r \simeq r_{dS}$
by setting the boundary condition $S|_{r = r_{dS} - 10^{-3}r_{H}} = 0$.
We can see that our method still works even in the almost marginally stable case.

\subsection{5-Dimensional Schwarzschild black string}
\label{5dblackstring}

\label{5dstring}
\begin{figure}
\includegraphics[width=8cm]{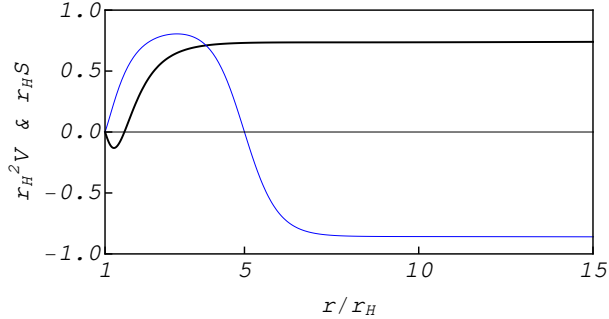}
\caption{\label{figbsting}  The effective potential $V$
of the scalar perturbation in the 5-Dimensional Schwarzschild black string (thick black)
with the parameter $r_H k = 0.877$,
and the bounded solution $S$ with the boundary condition $S|_{r = 5 r_{H}} = 0$ (solid blue).
 }
\end{figure}
Finally, we discuss the case of the 5-Dimensional Schwarzschild black string spacetime.
This spacetime has an unstable mode (Gregory-Laflamme instability) 
for a long wave perturbation $k < k_{\rm cr} \simeq 0.876$~\cite{Gregory:1993vy}.
The master equation for the scalar perturbation
also takes the Sch\"odinger form Eq.\eqref{schrodingereq} as shown in~\cite{Hovdebo:2006jy, Konoplya:2008yy}.
We should note that the effective potential Eq.\eqref{5dbspotential} is 
positive in $r > r_H$ if $k>(1+\sqrt{3})/\sqrt{2} \simeq 1.93$.
In Fig.\ref{figbsting},
we plot the effective potential and the numerical solution of Eq.\eqref{sdefvzeromain}
for $r_H k = 0.877$ with the boundary condition $S|_{r = 5 r_{H}} = 0$.
In this case, the value of the potential asymptotes to a positive constant at large distance.
{}From the same discussion in the proof of the proposition 1.(see Appendix.\ref{proof1}),
it is guaranteed that 
the solution is continous and bounded in $r>r_{\rm ini} = 5 r_H$.
Also, our method works for the almost marginally stable mode.

\section{summary and discussion}
\label{summary}

We have studied a
sufficient condition to prove the stability of 
a black hole when the linear perturbative equation 
takes the Schr\"odinger form by showing the 
$S$-deformation such that the deformed potential vanishes everywhere.
While our method is just a sufficient condition for the non-existence of bound state $\omega^2 < 0$,
it also becomes a necessary condition in a simple toy model.
We have also found the numerical solution $S$
for the vector and scalar perturbation on 10-dimensional spherically symmetric black holes,
and the scalar perturbation on the 5-dimensional Schwarzschild black string.
While $S$ diverges at some point for an inappropriate boundary condition, 
we found the continuous range of appropriate boundary conditions, 
which corresponds to bounded $S$.
Furthermore, as shown in Sec.\ref{10RNbh} and Sec.\ref{5dblackstring}, our method can work even in 
almost marginally stable cases.
From these results, we conjecture the following:
\\\\
{\bf Conjecture 1.}~~{\it A continuous and bounded solution of Eq.\eqref{sdefvzeromain}
exists in $-\infty < x < \infty$
if and only if there is no bound state ($\omega^2 < 0$ mode) in the Schr\"odinger equation. 
}
\\\\
At least, the present results suggest that our method is a good test for the stability of black hole.
As for the existence of the solution, 
we gave a proof for the positive and bounded potential which corresponds to a manifestly stable case.
If the potential $V$ contains negative regions, 
we can still show the existence 
from a continuous and bounded 
solution for a different potential $V_0 (\le V)$, 
under some assumptions (see Appendix.\ref{appendixc}).
The case of marginally (un)stable parameter was discussed in Sec.\ref{marginalystable}.

If there exists a regular solution of Eq.\eqref{sdefvzeromain}, 
the Schr\"odinger equation~\eqref{schrodingereq} becomes 
\begin{align}
\left(-\frac{d}{dx} + S \right)\left(\frac{d}{dx} + S\right)\Phi = \omega^2 \Phi.
\end{align}
This is known as the supersymmetric quantum mechanics system where the energy $\omega^2$ is 
manifestly non-negative.
If the above conjecture is correct, 
the quantum mechanics system which does not have a bound state with negative energy
becomes supersymmetric.

\section*{Acknowledgments}
The author would like to thank
Vitor Cardoso, Tsuyoshi Houri, George Pappas, Jiro Soda and Kentaro Tatsumi for their useful comments.
M.K. acknowledges financial support provided under the European Union's H2020 ERC 
Consolidator Grant ``Matter and strong-field gravity: New frontiers in Einstein's theory''
grant agreement no. MaGRaTh-646597, and under the H2020-MSCA-RISE-2015 Grant No. StronGrHEP-690904.

\appendix

\section{Proof of the proposition 1}
\label{proof1}
We give a proof of the proposition 1. in Sec.\ref{positivepotentialcase}.
\\
{\bf Proof}.~~
We consider to solve Eq.\eqref{sdefvzeromain} from the boundary condition $S|_{x = x_0} = 0$ at a point 
$x = x_0$.
We already know the local existence of the Eq.\eqref{sdefvzeromain}, so we only need to 
exclude possibility that $S$ is divergent at some point.\footnote{
In fact, we also need to exclude the possibility that 
$S$ is bounded above and below but it oscillates infinitely many times near some point $x = x_s$
and does not have a limiting point like $\sin(1/(x - x_s))$.
However, this does not happen, because if
$S$ oscillates infinitely many times in a finite interval, it's 
derivative should be unbounded above and below. 
In that case $dS/dx - S^2$ also becomes unbounded above and below, but this contradicts with
that $V$ is bounded above.} 
At $x = x_0$, we have $S|_{x = x_0} = 0$ and $dS/dx|_{x = x_0} = - V|_{x = x_0} < 0$,
so $S$ takes positive value at $x_0 - \delta_0 < x < x_0$, and negative value at $x_0 < x < x+\delta_0$ 
for a small constant $\delta_0 > 0$.

First, we consider the region $x_0 < x$.
We can say that once $S$ becomes negative, $S$ cannot be zero as $x$ increases in $x_0 < x$.
If $S = 0$ at some point $x_1 (> x_0)$ and $S < 0$ in $ x_0 < x < x_1$, 
then $dS/dx|_{x = x_1} = - V|_{x = x_1} < 0$.
However, this is a contradiction, because $dS/dx|_{x = x_1} < 0$ implies 
that $S$ is already positive in $x_1 - \delta_1 < x < x_1$ for a small constant $\delta_1 >0$. 
Thus, $S$ is bounded above in the region $x_0 < x$.

We also have $dS/dx = -V + S^2 > - (V_{\rm max} + \delta_2) + S^2$ for a small constant $\delta_2 >0$.
If the value of $S$ changes from the value larger than $- \sqrt{V_{\rm max} + \delta_2}$
into the value $- \sqrt{V_{\rm max} + \delta_2}$ at $x = x_2 (> x_0)$, 
$dS/dx$ becomes positive at $x = x_2$. However this is a contradiction because,
$S|_{x = x_2} = - \sqrt{V_{\rm max} + \delta_2}$ and $dS/dx|_{x =x_2} > 0$ implies 
$S > - \sqrt{V_{\rm max} + \delta_2}$ in $x_2 > x > x_2 - \delta_3$ for a small constant $\delta_3 >0$.
Since $\delta_2$ is an arbitrary constant, $S$ cannot be smaller than $- \sqrt{V_{\rm max}}$.
This shows that $S$ is bounded below in the region $x_0 < x$.

Next, we consider the region $x < x_0$.
Defining $\bar{S} := - S, \bar{x} := - x$, then Eq.\eqref{sdefvzeromain} becomes
\begin{align}
\frac{d\bar{S}}{d\bar{x}} = - V + \bar{S}^2,
\end{align}
then $x = x_0$ corresponds to $\bar{x} = - x_0 =: \bar{x}_0$.
We need to show $\bar{S}$ is bounded above and below in $ \bar{x}_0 < \bar{x}$,
but this is formally the same problem as in the case of $x_0 < x$.
Thus, $S$ is bounded above and below in the region $-\infty < x < \infty$.
\hfill$\Box$

\section{Effective potential for $D$-dimensional Reissner-Nordstr\"om-de Sitter black hole}
\label{appendixa}

The metric for the $D$-dimensional Reissner-Nordstr\"om-de Sitter black hole is
\begin{align}
ds^2 &= - f dt^2 + \frac{dr^2}{f} + r^2 d\Omega_{S^n},
\\
f &= 1 - \frac{2M}{r^{n-1}} + \frac{Q^2}{r^{2n-2}} - \lambda r^2,
\end{align}
where $n = D-2$, $\lambda = 2\Lambda/(n(n+1))$ with a cosmological constant $\Lambda$,
and $d\Omega_{S^n}$ denotes the metric of the unit $n$-dimensional sphere.
As shown in~\cite{Kodama:2003jz, Ishibashi:2003ap, Kodama:2003kk}, 
the linear gravitational and electromagnetic perturbation reduces to 
the decoupled single master equations 
with the same form as Eq.\eqref{mastereqtr}.
The effective potentials for the vector and scalar gravitational modes~\footnote{
The effective potentials for other modes, tensor perturbation and 
electromagnetic vector and scalar perturbation, 
can be seen in~\cite{Kodama:2003jz, Ishibashi:2003ap, Kodama:2003kk}.} 
are
\begin{align}
V_{\rm vector} &= \frac{f}{r^2} \bigg[ \ell(\ell + n -1)
+
\frac{n^2- 2n }{4}
-
\frac{n(n-2)}{4} \lambda r^2
+
\frac{n(5n-2)Q^2}{4 r^{2n-2}}
\notag\\& \quad
+
\frac{1}{r^{n-1}}
\bigg(
-\frac{n^2 + 2}{2}M 
- 
\sqrt{(n^2 - 1)^2 M^2 + 2n(n-1)(\ell +n)(\ell -1)Q^2}
\bigg)
\bigg],
\\
V_{{\rm scalar}} &= \frac{f}{64 r^2} \frac{U}{H^2},
\end{align}
with
\begin{align}
U &= 
\Big[-4 n^3 (n + 2) (n + 1)^2 (1 + m \delta)^2 X^2 + 
   48 n^2 (n + 1) (n - 2) m (1 + m \delta) X 
\notag \\ & \quad - 
   16 (n - 2) (n - 4) m^2\Big] \lambda r^2
-n^3 (3 n - 2) (n + 1)^4 \delta (1 + m \delta)^3 X^4
\notag\\ & \quad
-4 n^2 (n + 1)^2 (1 + m \delta)^2 \Big[(n + 1) (3 n - 2) m \delta - 
    n^2\Big] X^3
\notag\\ & \quad
+4 (n + 1) (1 + 
   m \delta) \Big[m (n - 2) (n - 4) (n + 1) (m + n^2) \delta + 
   4 n (2 n^2 - 3 n + 4) m 
\notag \\ & \quad 
+ n^2 (n - 2) (n - 4) (n + 1)\Big] X^2
-16 m ((n + 1) m (-4 m + 3 n^2 (n - 2)) \delta 
\notag \\ & \quad 
+ 3 n (n - 4) m 
+ 
   3 n^2 (n + 1) (n - 2)) X
+ 64 m^3 + 16 n (n + 2) m^2,
\\
H &= m + \frac{n(n+1)}{2}(1 + m \delta)X,
\\
X &= \frac{2M}{r^{n-1}},
\\
\delta &= \frac{\mu - M}{2 m M},~~\mu = \sqrt{M^2 + \frac{4 M Q^2}{(n+1)^2}},~~ m = \ell(\ell + n-1) - n,
\end{align}
where $\ell$ is a positive integer, $\ell \ge 1$ for vector mode, $\ell \ge 0$ for scalar mode.
The relation between $x$ and $r$
is $dx = dr/f(r)$. 
$x$ behaves $x \sim \ln(r-r_H)$ near the horizon.
In the coordinate $r$, Eq.\eqref{sdefvzeromain} becomes
\begin{align}
V + f\frac{dS}{dr} - S^2 = 0.
\end{align}

To normalize the all quantities by the radius of the black hole horizon $r_H$,
we set the mass parameter as
\begin{align}
M &= \frac{1}{2}\left(1 + Q^2 - \lambda
\right),
\end{align}
then the black hole horizon locates at $r = r_H = 1$.
Also, setting
\begin{align}
\lambda = 
\frac{1 - \rho^{1-n} + Q^2 (1-\rho^{n-1})}{1 - \rho^{-n-1}},
\end{align}
the location of the de-Sitter horizon becomes $r = r_{dS} = 1/\rho$.
The electric charge for the extremal black hole is given by
\begin{align}
Q^2_{\rm extremal} = \frac{(n-1)(1-\rho^{n+1})-\rho^2 (n+1)(1-\rho^{n-1})}{(n-1)(1-\rho^{n+1}) - \rho^{n+1}(n+1)(1-\rho^{n-1})},
\end{align}
These useful normalization was used in~\cite{Konoplya:2007jv}.

\section{Effective potential for 5-dimensional black string}
\label{appendixblackstring}

The metric of the 5-dimensional black string is
\begin{align}
ds^2 = -f dt^2 + \frac{dr^2}{f} + r^2 (d\theta^2 + \sin^2\theta d\phi^2) + dz^2,
\end{align}
with $f = 1 - r_H/r$.
The effective potential for the scalar perturbation is given by~\cite{Konoplya:2008yy}
\begin{align}
V = \frac{f}{r^3}
\frac{k^6 r^9 + 3 k^4 r^6 (2 r - 3 r_H) + r_H^3 + 
   3 k^2 r^3 r_H (-4 r + 3 r_H)}{(k^2 r^3 + r_H)^2},
\label{5dbspotential}
\end{align}
where $k$ is the wave number along $z$ direction.
In this case, Eq.\eqref{sdefvzeromain} also becomes
\begin{align}
V + f\frac{dS}{dr} - S^2 = 0.
\end{align}

\section{$V = - a^2 \exp(b^2 x)$ case}
\label{appendixb}

In some cases, the potential 
behaves $V \simeq - a^2 (r-r_H)$ with a constant $a >0$ near the horizon.
Since $r - r_H \simeq e^{b^2 x}$ with a constant $b>0$ near the horizon, 
the potential in $x$ coordinate becomes $V \simeq - a^2 \exp(b^2 x)$.
For this potential, we can solve Eq.\eqref{sdefvzeromain} by using the Bessel functions
\begin{align}
S &= b^2 X \frac{C_1 J_1(X) + 2  Y_1(X)}{2 C_1 J_0(X) + 4  Y_0(X)},
\end{align}
where $X = 2 a e^{b^2 x/2} / b^2$ and $C_1$ is a constant.
For $X \ll 1$, the Bessel functions behave
\begin{align}
J_0(X) &= 1 + {\cal O}(X^2),
\\
J_1(X) &= \frac{X}{2} + {\cal O}(X^2),
\\
Y_0(X) &= \frac{2}{\pi}\ln X
+
2\frac{\gamma - \ln 2}{\pi} 
 + {\cal O}(X^2),
\\
Y_1(X) &= - \frac{2}{\pi}\frac{1}{X} + X \frac{1}{\pi} \ln X
- \frac{1 - 2\gamma + 2 \ln 2}{2\pi}X + {\cal O}(X^2),
\end{align}
where $\gamma \simeq 0.577$ is the Euler's constant.
So, the approximate solution of $S$ with the integral constant becomes
\begin{align}
S \simeq  - \frac{b^2}{2} \frac{1}{C_2 + \ln X},
\end{align}
where $C_2 = \gamma - \ln 2 + C_1 \pi/4$.
Once $S$ takes positive value in $X \ll 1$ regime, 
$S$ will be bounded above and below as $X$ decreases towards zero.
In that case, near $X\simeq 0$, $S$ behaves $S \sim - 1/\ln X \sim -1/x$ like $V = 0$ case.

\section{A potential with compact support}
\label{appendixc}

Let us consider a continuous and bounded potential with compact support
\begin{align}
V = 
\begin{cases}
 0   &(-\infty < x \le x_1)
\\ 
 V_- ~( < 0)   &(x_1 < x < x_2)
\\
0 &(x = x_2)
\\
 V_+ ~( > 0) &(x_2 < x < x_3)
\\
  0  & (x_3 \le x < \infty).
\end{cases}
\label{generalpotentialV01}
\end{align}
We assume that $V$ is smooth in $x_1 < x < x_3$ and 
$\lim_{x \to x_3-0} dV/dx > 0$.
Also, we define another potential which is not necessary continuous
\begin{align}
V_0 = 
\begin{cases}
 0   &(-\infty < x \le x_1)
\\ 
v_- ~( v_- < V_- < 0)   &(x_1 < x < x_2)
\\
v_+ ~( 0 \le v_+ < V_+ ) &(x_2 \le x < x_3)
\\
  0  & (x_3 \le x < \infty),
\end{cases}
\label{generalpotentialv01}
\end{align}
where $v_\pm$ are functions 
and we assume $V > V_0$ in $x_1 < x < x_3$. In this set up, we would like to prove the following proposition.
\\\\
{\bf Proposition 2.}~~{\it If there exists a continuous and bounded solution of Eq.\eqref{sdefvzeromain} for the potential Eq.\eqref{generalpotentialv01}
in $-\infty < x < \infty$, 
there also exists a continuous and bounded solution of Eq.\eqref{sdefvzeromain} for the potential Eq.\eqref{generalpotentialV01} in $-\infty < x < \infty$.}
\\\\
{\bf Proof}.~~
Let $S_0$ be a continuous and bounded solution of $V_0 + dS_0/dx - S_0^2= 0$.
Since the behavior in $x \le x_1$ and $x_3 \le x$ are the same as the toy model in Sec.\ref{toymodel},
we can say $S_0|_{x = x_1} > 0$ and $S_0|_{x = x_3} < 0$.
In the region $x_1 < x < x_2$, since $-V_- < -v_-$, we obtain an inequality
\begin{align}
(S_0  - S)(S_0 - S + 2 S) <  \frac{d(S_0 - S)}{dx}. 
\end{align} 
We assume $S|_{x = x_1} = \eta S_0|_{x = x_1}$ with a positive constant $\eta (< 1)$.\footnote{Note that this is an appropriate boundary condition for $S$.}
Since $dS/dx = -V_- + S^2 > 0$, $S$ is a a monotonically increasing function in $x_1 < x < x_2$.
{}From this and the condition $S|_{x = x_1} = \eta S_0 > 0$, we can say $S > 0$ in $x_1 < x < x_2$.
In the limit $x \to x_1 +0$, the above inequality becomes
\begin{align}
(1-\eta^2)S_0^2 < \frac{d(S_0 - S)}{dx},
\end{align}
and LHS is positive. 
Thus $S_0 - S$ is an increasing function near $x = x_1$.
Once $S_0 - S$ becomes positive, it cannot be zero in $x_1 < x < x_2$
because $d(S_0 - S)/dx > 0$ for positive $S_0 - S$ in $x_1 < x < x_2$, {\it i.e.}, 
$S_0 - S$ cannot decrease as $x$ increases.
So, we can say $S < S_0 $ in $x_1 < x < x_2$.
Since $S_0$ is bounded above, $S$ cannot be divergent in $x_1 < x < x_2$.

In the region $x_2 < x < x_3$, since $-V_+ < - v_+$ we also have the same inequality
\begin{align}
(S_0  - S)(S_0 - S + 2 S) <  \frac{d(S_0 - S)}{dx}. 
\end{align} 
If we assume $S \ge 0$ in $x_2 < r < x_3$, we can also say $S < S_0$ from the same 
discussion above.
However, now we also assumed $S_0 <0$ 
at some point because $S_0$ should be connected with $1/(c_4 -r)$, which is negative, at $x = x_3$,
then this contradicts with the first assumption $S>0$.
Thus, we can say $S < 0$ at some point $x = y_0 $ which satisfies $x_2 < y_0 < x_3$. 
In $y_0 < x < x_3$, 
from the same discussion in the proof of the proposition 1,
we can show that $S$ is a continuous and bounded negative function.

There is still a possibility that $S = 0$ at $x = x_3$ where $V_+ = 0$.
In that case, $S|_{x= x_3} = 0, dS/dx|_{x=x_3} = 0$, and
$d^2 S/dx^2|_{x = x_3} = \lim_{x \to x_3-0} dV/dx (=: \beta^2) > 0$. 
So, the approximate behavior becomes $S \simeq \beta^2(x -x_3)^2$,
but this means $S > 0$ in $x_3 - \delta < x < x_3$ with a small constant $\delta >0$.
This contradicts with $S < 0$ in $y_0 < x < x_3$.
Thus, $S|_{x = x_3} < 0$ and $S$ can be matched with $1/(c_4 -r)$ at $x = x_3$ with a negative value.
This shows the existence of continuous and bounded $S$ in $-\infty < x <\infty$.
\hfill$\Box$

\end{document}